\documentclass[doublecol,figures]{epl2}
\usepackage{epsfig}
\usepackage{amsmath}
\usepackage{color}
\title{Calculation of the even-odd energy difference in superfluid Fermi systems using the pseudopotential theory}
\shorttitle{Calculation of the even-odd energy difference in superfluid Fermi systems etc.}
\author{A. Csord\'as\inst{1,2}\thanks{E-mail: \email{csordas@tristan.elte.hu}} 
\shortauthor{A. Csord\'as et al.}
\and G. Homa\inst{1}\thanks{E-mail: \email{ggg.maxwell1@gmail.com}} \and P. Sz\'epfalusy\inst{1,3}\thanks{E-mail: \email{psz@complex.elte.hu}}}
\institute{
\inst{1} Department of Physics of Complex Systems, E\"otv\"os University, P\'azm\'any P. S\'et\'any 1/A, H-1117 Budapest, Hungary, EU \\
\inst{2} HAS-ELTE Statistical and Biological Physics Research Group, E\"otv\"os University, P\'azm\'any P. S\'et\'any 1/A, H-1117 Budapest, Hungary, EU \\  
\inst{3} Research Institute for Solid State Physics and Optics,
             P. O. Box 49, H-1525 Budapest, Hungary, EU
}
\date{\today}
\pacs{74.20.Fg}{BCS theory and its development}
\pacs{67.85.Lm}{Degenerate Fermi gases}
\pacs{71.15.Dx}{Computational methodology (Brillouin zone sampling, iterative diagonalization, pseudopotential construction)}

\abstract{
The pseudopotential theory is extended to the Bogoliubov-de Gennes equations to determine the excess energy when one atom is added to the trapped superfluid Fermi system with even number of atoms. Particular attention is paid to systems being at the Feshbach resonance point. The results for relatively small particle numbers are in harmony with the Monte Carlo calculations, but are also relevant for systems with larger particle numbers.
Concerning the additional one quasiparticle state we define and determine two new universal numbers to characterize its widths.}
\begin{document}
\maketitle


The pseudopotential theory has proved to be an important tool in
different areas, as for instance solid state physics (see,
e.g., \cite{Abrikosov1988}), quantum chemistry (see,
e.g., \cite{Szasz1985}), metal clusters (see,
e.g., \cite{Lipparini2003}). A general survey of the method regarding
the interest of researchers in a wide range of applications is given
in \cite{Payne1992}. 
For a recent review about the approximations in
electronic structure energy see ref.\cite{Schwerdtfeger2011}. 
The idea was first formulated by Hellmann \cite{Hellmann1935} and
appeared also in Gomb\'as' work \cite{Gombas1935}. The quantum
mechanical foundation was started by F\'enyes \cite{Fenyes1943} (for the early history of the pseudopotential theory see  
\cite{Gombas1967,Schwarz1968}). 

In the quantum mechanical treatment of many particle systems it is
crucial to reduce the task from an $N_{total}$-particle problem to
an $N_{eff}$ particle one, where $N_{eff} \ll N_{total}$, when
$N_{total} \gg 1$. In atoms, molecules and metals, e.g., $N_{eff}$ is
the number of valence electrons. The pseudopotential method works by
taking the advantage of the fact that the large negative potential
energy felt by a valence electron is almost completely cancelled by
its large positive kinetic energy, which is due to the oscillations of
its wave function inside the core. The method deals with pseudo wave
functions which are smooth and easier to handle. One expects that
similar situation arises in any fermion system put in an external
potential when the analog of the valence electron problem in a broad
sense is investigated. Characteristic differences also show up in case
of trapped Fermi gases as a consequence of the short-range nature of
the interaction between the atoms and in particular when the
superfluid state is realized, especially at the Feshbach resonance.

The main goal of the present paper is to formulate and use the pseudopotential method for trapped Fermi gases at zero temperature. For the sake of concreteness the internal degree of freedom of the atoms will be assumed to be twofold, one can have in mind for instance the ${}^6$Li isotopes as elements of a two component Fermi gas. The BCS side of the BCS-BEC crossover will be considered, where the interparticle interaction can be characterized by a negative $s$-wave scattering length $a$. We will focus in particular on the Feshbach resonance point, where universal properties set in, since $|a| \to \infty$ there, and the range of the interaction can be neglected (for a review see \cite{Giorgini2008}).

We start by recalling the Bogoliubov-de Gennes (BdG) equations, which can be considered as the generalization of the Hartree-Fock equations to include pair correlations and conceived as equations of a mean-field theory \cite{Ring2005,deGennes1966}. They are as follows:
\begin{eqnarray}
\left[ H+U_{int} \right]u_n +\Delta v_n &=& \epsilon_n u_n, \nonumber\\
\Delta^*u_n- \left[H +U_{int} \right]v_n &=& \epsilon_n v_n, \label{eq:BDE}
\end{eqnarray}
or in short
\begin{equation}
\hat\Omega \phi_n=\epsilon_n \phi_n 
\label{eq:shortBDE}
\end{equation}
with $\phi_n=\left( u_n, v_n \right)$. The operator $H$ can be written as
\begin{equation}
H= -\frac{\hbar^2}{2m} \nabla^2  + U_{ext}(\mathbf{r})-\mu,
\end{equation}
where $m$ stands for the mass of the atom, $\mu$ for the chemical
potential and $U_{ext}$ for the confining potential, which will be
assumed to be harmonic in the applications. $U_{int}(\mathbf{r})$
denotes the diagonal part of the mean-field potential, while
$\Delta(\mathbf{r})$ is the pair potential of the mean field
theory. In a more general framework both $U_{int}(\mathbf{r})$ and
$\Delta(\mathbf{r})$ contain contributions from correlation effects,
but in the local density approximation (LDA), adopted also here, they
are local quantities. We are interested in a system of particle number
$N+1$ with $N$ being even. In particular we want to calculate the
change in the ground state energy when adding one particle to the $N$-particle system. Since the theory adopted here is not working with
fixed particle numbers it is more precise to speak about even- and odd-number parity states \cite{Ring2005}, but for sake of simplicty the
therminology above is often used. 

It is assumed in the following that the $N$-body problem is
solved. It might mean the solution of the BdG equations in the 
BCS mean-field approach, in which case $U_{int}(\mathbf{r})$ is
generally put zero \cite{Giorgini2008,Antezza2007} and
only $\Delta(\mathbf{r})$ should be solved self-consistently. It is
argued that in the mean-field description this is a 
consistent choice \cite{Liu2007}. At unitarity, however, there
exists an extension 
of the density functional theory, which determines also $U_{int}(\mathbf{r})$
self-consistently \cite{Bulgac2007,Bulgac2010,Bulgac2011}. In our
general considerations 
there is no need to 
specify which case is adopted. We do not use the results
explicitely, suppose only their existence in principle
and do not restrict ourselves to such particle numbers for which
calculations have been realized. 
We avoid the details of the direct interaction $U_{int}$ by
fixing $U_{int}$ and $\Delta$ using the solution of the $N$-body
problem. 
In a more precise treatment one
ought to solve the $(N+1)$-particle problem self-consistently. (In a
normal system, as the electrongas in an 
atom for instance, this amounts to
neglecting the core polarization by the valence electron.) In our
model pseudopotential at unitarity universal constants will carry all
the informations concerning the $N$-atom problem. Therefore, we do not
need to specify a regularisation procedure, as in 
refs.~\cite{Antezza2007,Bulgac2011}.

It will be important in the following that eqs. (\ref{eq:BDE}) provide
positive and negative eigenvalues as well representing the charge
conjugation symmetry \cite{Bulgac2010,Blaizot1986}. More
concretely if 
$\phi^{(+)}=\left(u, v \right)$ is an
eigenfunction with eigenvalue $ \epsilon > 0 $ then
$\phi^{(-)}=\left(-v^* , u^* \right)$
is an eigenfunction with eigenvalue $-\epsilon$. The
orthonormalization condition 
\begin{equation}
\int d^3 r \, \left[ u_n^* u_m+ v_n^* v_m\right]=\delta_{n,m}
\label{eq:normalization}
\end{equation}
leads to the conclusion that eigenfunctions belonging to eigenvalues
of different sign are orthogonal.

Let us denote by $E^{(N)}$ the ground state energy of the $N$ particle
system, then, see, e.g., ref.~ \cite{Ring2005}
\begin{equation}
E^{(N+1)}-E^{(N)}=\epsilon_{min}+\mu, \quad N  \mbox{ is even}.
\label{eq:energy_change}
\end{equation}
For $\mu$ the chemical potential of the $N$-body system should be
taken.  
The calculation of $\epsilon_{min}$ as the smallest
positive eigenvalue of eq.~(\ref{eq:shortBDE}) raises severe technical
difficulties for large $N$, since the corresponding
eigenfunction becomes highly oscillating.
To avoid it and to make the determination of $\epsilon_{min}$ manageable even in such a case we generalize the pseudopotential method to the BdG equations by introducing 
the modified equation 
\begin{equation}
\left( \hat\Omega+\hat{U}_{ps}\right)\tilde{\phi}_n=\epsilon_n\tilde{\phi}_n,
\label{eq:modBDE}
\end{equation}
where $\tilde{\phi}_n$ stands for the pseudo wave function 
$\tilde{\phi}_n(\mathbf{r})=\left( 
\tilde{u}_n(\mathbf{r}),  \tilde{v}_n(\mathbf{r})
\right)$, 
and $\hat{U}_{ps}$ is the nonlocal pseudopotential
\begin{eqnarray}
\hat{U}_{ps}&=&\sum_t \left\{ \left( \epsilon_n-\epsilon_t\right)
\left( \begin{array}{c} u_t(\mathbf{r}) \\ v_t(\mathbf{r}) \end{array} \right)
\left(u_t^*(\mathbf{r'}),v_t^*(\mathbf{r'})\right)\right.
\nonumber \\
&&+\left.\left( \epsilon_n+\epsilon_t\right)
\left( \begin{array}{c} -v_t^*(\mathbf{r}) \\
 u_t^*(\mathbf{r}) \end{array}
\right)\left(-v_t(\mathbf{r'}),u_t(\mathbf{r'})\right) 
\right\}. \nonumber \\
\label{eq:nonlocal_pot}
\end{eqnarray}
Here $\epsilon_t$ is positive by
definition. $\phi_t(\mathbf{r})=(u_t(\mathbf{r}),v_t(\mathbf{r}))$'s
are selected of those solutions of eq.~(\ref{eq:BDE}) that are
involved in building the functions $U_{int}(\mathbf{r})$ and
$\Delta(\mathbf{r})$.

Two remarks are in order here. One has to include states of positive and negative eigenvalues to maintain the "charge conjugation" symmetry. 
The summation may extend only to a restricted set to avoid divergences, for instance to states 
\begin{equation}
\left[\langle u_t | H+U_{int} | u_t \rangle +\langle v_t | H+U_{int} | v_t \rangle\right]<0. 
\label{eq:crit}
\end{equation}
This choice will be assumed in the following. 
One can easily see that the solution of eq. (\ref{eq:modBDE}) is not unique, any linear combination of $\phi_t^{(+)}$ and $\phi_t^{(-)}$ (satisfying the condition (\ref{eq:crit})) can be added to 
$\tilde{\phi}_n$ without changing the eigenvalue $\epsilon_n$. As a consequence the most general form of $U_{ps}$ can be written as
\begin{equation}
U_{ps}=\sum_t \left\{ \left| \phi_t^{(+)}\right\rangle \left\langle F^{(+)}_t\right|+ \left| \phi_t^{(-)}\right\rangle \left\langle F^{(-)}_t\right|\right\}.
\label{eq:upsgener}
\end{equation}
One can easily convince oneself that
\begin{equation}
\left\langle F^{(\pm)}_t\right|\left.\tilde{\phi}_n\right\rangle =
\left( \epsilon_n \pm \epsilon_t \right)
\left\langle \phi^{(\pm)}_t\right|\left.\tilde{\phi}_n\right\rangle;
\label{eq:ftchoice}
\end{equation}
$F^{(\pm)}$ are quite general functions. This expression of the pseudopotential is the generalization of the Austin, Heine Sham proposal \cite{Austin1962}. A natural requirement might be that the pseudo wave function be as smooth as possible (formally, it has been proposed to minimize the kinetic energy 
\cite{Szepfalusy1956,Cohen1961}, see also
\cite{Szasz1985}). An obvious condition is that the pseudopotential should cancel a large part of the external and the local mean-field potentials (see e.g., \cite{Hellmann1937,Abrikosov1988,Szasz1985}).

In this spirit we choose the generalized Austin type \cite{Austin1962,Abrikosov1988} form
\begin{equation}
\hat{U}_{ps}=\left(\begin{array}{cc}
U_{ps}^H & U_{ps}^\Delta \\
U_{ps}^{\Delta *} & -U_{ps}^{H *}  
\end{array}\right)
\label{eq:upsmatrix}
\end{equation}
with
\begin{eqnarray}
U_{ps}^H &=&-{\sum_t}' \Bigl( u_t(\mathbf{r})\left[ U(\mathbf{r}')-\mu \right]u_t^*(\mathbf{r}')\Bigr. \nonumber\\
 &&\phantom{-{\sum_t}'}\Bigl. +v_t^*(\mathbf{r})\left[ U(\mathbf{r}')-\mu \right]v_t(\mathbf{r}')\Bigr),
\label{eq:upsh}
\end{eqnarray}
\begin{eqnarray}
U_{ps}^\Delta &=&-{\sum_t}' \Bigl( v_t(\mathbf{r})\left[ U(\mathbf{r}')-\mu \right]u_t^*(\mathbf{r}')\Bigr. \nonumber\\
 &&\phantom{-{\sum_t}'}\Bigl. -u_t^*(\mathbf{r})\left[ U(\mathbf{r}')-\mu \right]v_t(\mathbf{r}')\Bigr).
\label{eq:upsdelta}
\end{eqnarray}
The prime on $\sum$ means that the summation extends to states specified above and 
\begin{equation}
U=U_{ext}+U_{int}.
\end{equation} 
This expression of the pseudopotential corresponds to the choice
\begin{eqnarray}
F_t^{(+)}(\mathbf{r})&=&-\left(U(\mathbf{r})-\mu\right)
\biggl( u_t(\mathbf{r}),v_t(\mathbf{r})\biggr), \\
F_t^{(-)}(\mathbf{r})&=&-\left(U(\mathbf{r})-\mu\right)
\biggl( -v_t^*(\mathbf{r}),u_t^*(\mathbf{r})\biggr).
\end{eqnarray}

We consider in the following a trapped gas at the Feshbach resonance,
where universal properties characterize the system. Let's define the
Thomas-Fermi region (TFR), where the conditions for the Thomas-Fermi
approach are fulfilled (here and in the following, of course, the
generalized Thomas-Fermi theory including paircorrelations is meant
supposing that the particle number is large enough
\cite{Szepfalusy1964,Csordas2010}). An essential contribution to
${\sum_t}'\left(u_t(\mathbf{r})u_t^*(\mathbf{r}')+
  v_t(\mathbf{r})v_t^*(\mathbf{r}')\right)$ arises only in the TFR
then and it can be taken zero outside this region. Furthermore,  
it can be replaced to a good approximation, by a Dirac-delta function
$\delta(\mathbf{r}-\mathbf{r}')$ within the TFR if the number of
particles is large enough, which in other words means that the
nonlocality of the pseudopotential is disregarded. Concerning 
(\ref{eq:upsdelta}) it is decisive that in LDA $u_t$  and $v_t$ are
plane waves and the condition (\ref{eq:crit}) selects wave numbers in
pairs $\mathbf{k},-\mathbf{k}$, that altogether leads to $U_{ps}^\Delta=0$.
The pseudopotential becomes diagonal in this approximation 
\begin{equation}
\hat{U}_{ps}=\begin{cases}  \left( \begin{array}{cc} 
\mu - U(\mathbf{r}) & 0 \\
0 & U(\mathbf{r})-\mu 
\end{array} \right) \delta(\mathbf{r}-\mathbf{r}'), & \mathbf{r} \in \textrm{TFR} \\
 \quad \mathbf{0}, &  \textrm{otherwise.}  
\end{cases}
\label{eq:upsdiagonal}
\end{equation}
This pseudopotential can be regarded as a model one, since the approximation breaks down near the border of the TFR. The total hamiltonian of the model reads as 
\begin{equation}
\hat\Omega+\hat{U}_{ps}=\left(\begin{array}{cc}
-\frac{\hbar^2 \nabla^2}{2m} & \Delta(\mathbf{r}) \\
 \Delta(\mathbf{r}) & \frac{\hbar^2 \nabla^2}{2m}
\end{array}\right),\quad   \mathbf{r} \in \textrm{TFR},
\label{eq:tothamrtf}
\end{equation}
and in the region $\mathbf{r}\not\in\textrm{TFR}$:
\begin{equation}
\hat\Omega+\hat{U}_{ps}=\left(\begin{array}{cc}
-\frac{\hbar^2 \nabla^2}{2m} + U_{ext}-\mu & 0 \\
0 & \frac{\hbar^2 \nabla^2}{2m} - U_{ext}+\mu 
\end{array}\right).
\label{eq:tothamnotrtf}
\end{equation}
Here we used the fact that $U_{int}=0$ and  $\Delta(\mathbf{r})=0$ in LDA outside the TFR. 

At unitarity
\begin{equation}
\Delta(\mathbf{r})=\delta \left(\mu-U_{ext}(\mathbf{r})\right),
\label{eq:feshbachrel}
\end{equation}
where $\delta $ is a universal constant (see,
e.g., \cite{Giorgini2008}). 
In our calculations $\delta$ is an input. In contrast,
$\delta$ is obtained as an output from the methods used in
refs.~\cite{Antezza2007,Bulgac2011}.

\begin{figure}
\centering{
\epsfig{file=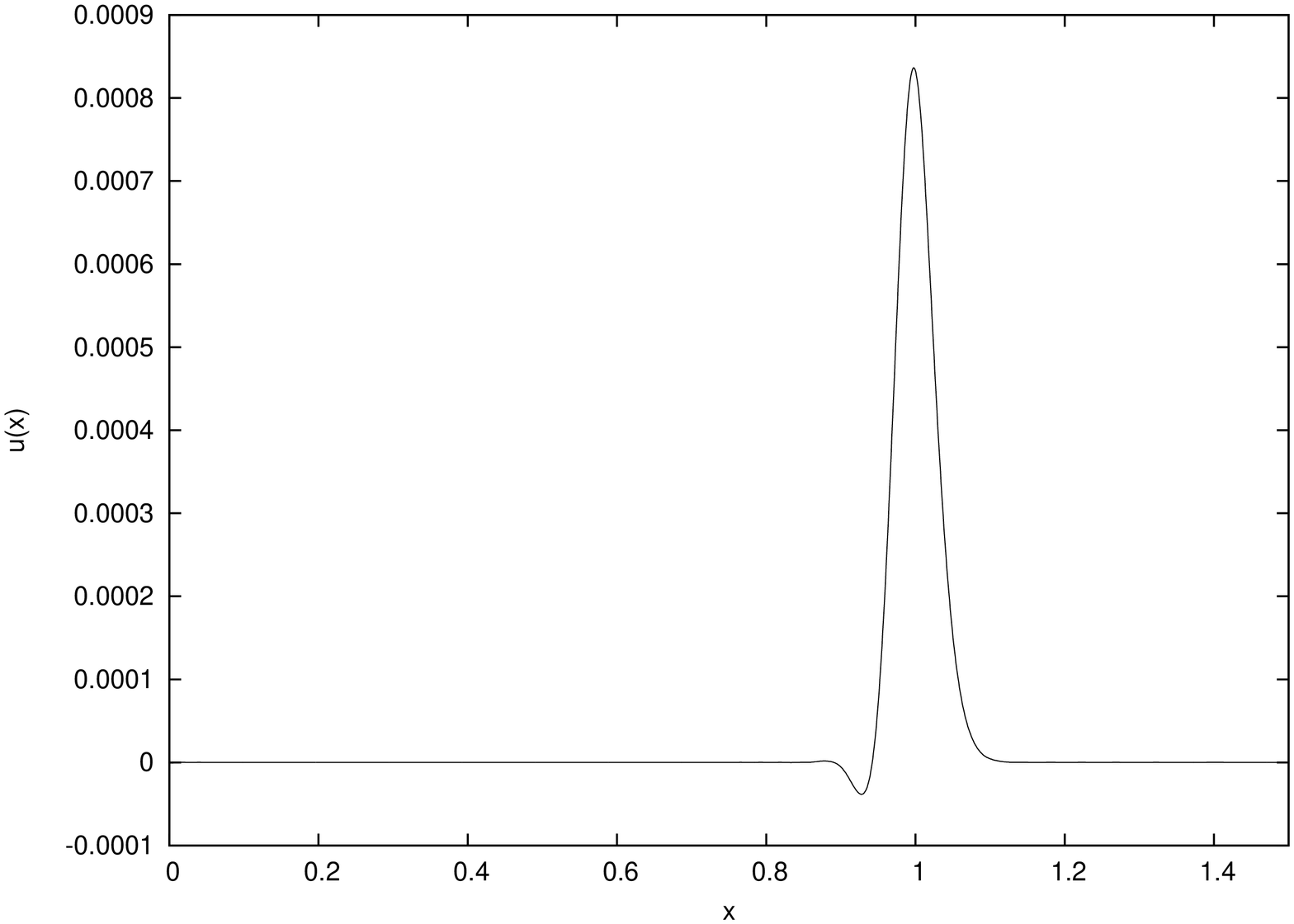,height=0.45\textwidth,angle=270}}
\centering{
\epsfig{file=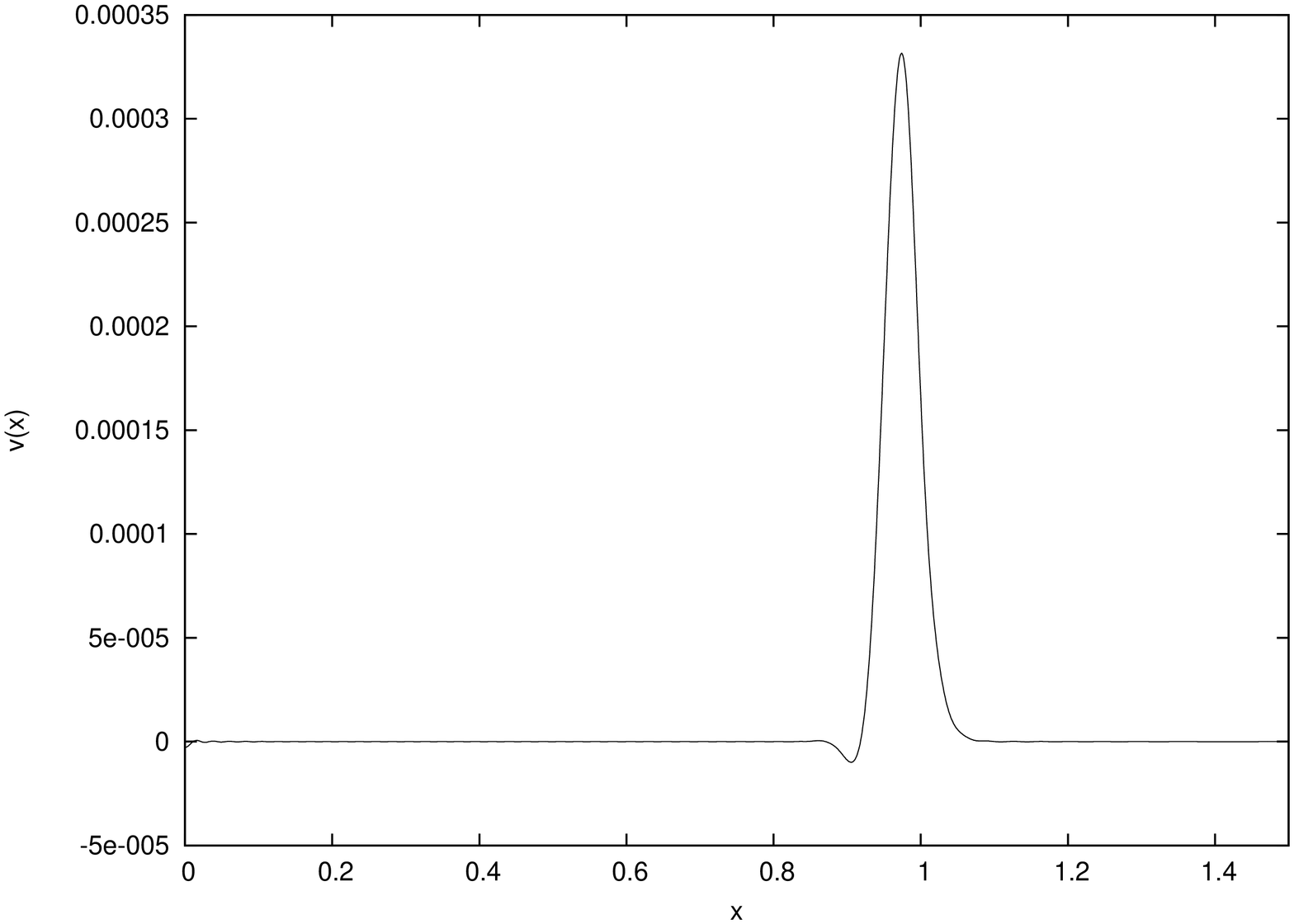,height=0.45\textwidth,angle=270}}
\caption{The $\tilde{u}$ (upper part) and the $\tilde{v}$ (lower part) components of the pseudo wave function as a function of the dimensionless variable of $x=r/R$. Parameters: $\mu=100\, \hbar\omega$, $\epsilon_{min}=4.437 \,\hbar\omega$, $\delta=1.16$.
\label{fig:uv}}
\end{figure}

\begin{figure}
\epsfig{file=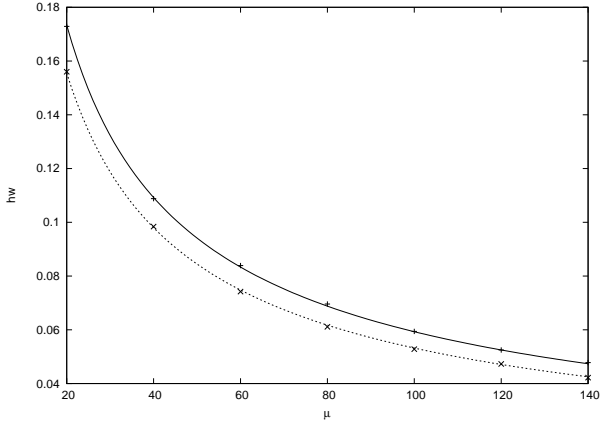,height=0.45\textwidth,angle=270}
\caption{The width of $\tilde{u}$ ($+$) and $\tilde{v}$ ($\times $) belonging to the pseudo wave function for $\epsilon_{min}$ as a function of $\mu$. The widths are measured in unit of $R$ and $\mu$ in $\hbar\omega$. The solid and dashed lines show the $w_u$ and $w_v$ curves (see text). 
\label{fig:widthscal}}
\end{figure}

We have carried out the calculation for a spherical symmetric harmonic oscillator trap potential 
\begin{equation}
U_{ext}(\mathbf{r})=\frac{1}{2}m\omega^2 r^2\equiv\mu x^2,
\end{equation}
where $x=r/R$ with $R$ being the Thomas-Fermi radius. 
First we have looked for a spherically symmetric solution (i.e., $l=0$)
to get the smallest eigenvalue of eqs.~(\ref{eq:modBDE}),(\ref{eq:tothamrtf}),(\ref{eq:tothamnotrtf}). 
\begin{figure}
\epsfig{file=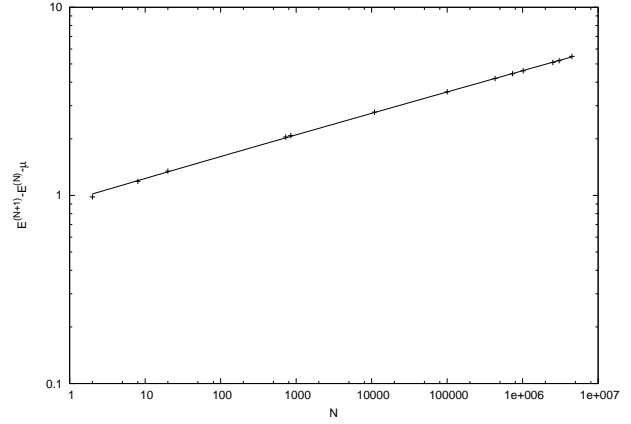,height=0.45\textwidth,angle=270}
\caption{Energy difference $E^{(N+1)}-E^{(N)}-\mu$ in units of $\hbar\omega$ as a function of $N$. Both axes are logarithmic, $\delta=1.16$. Straight line shows an $N^{1/9}$ behavior.
\label{fig:energydiff}}
\end{figure}
In principle one can solve the BdG equations for
$\mathbf{r}\in \mbox{TFR}$ and $\mathbf{r} \not\in \mbox{TFR}$ by
supposing that $\tilde{\phi}_n$ is finite and vanishes as
$\mathbf{r} \to \infty$, and by carefully matching the wave functions
at the border of TFR. But from the numerical point of view it is more
controllable the method of expansion in a given basis set. In the
numerics we expand both components of $\tilde{\phi}_n$ in the 3D
harmonic oscillator basis $\phi_{nlm}(\mathbf{r})$. In the isotropic
case $l$ and $m$ are good quantum numbers. Thus, the expansion is over
the different $n$ (radial quantum  number) values.
Matrix elements of
the kinetic energy and $U_{ext}$ in this basis are known
exacly. Numeric integration is applied for the matrix elements of
$\Delta(\mathbf{r})$ and $\hat{U}_{ps}$ (which are nonzero for
$\mathbf{r} \in TFR$). After calculating the matrix elements of
$\hat{\Omega} +\hat{U}_{ps}$ a simple diagonalization yields the
eigenvalue $\epsilon_n$ and the expansion coefficients of $u_n$ and
$v_n$. Special attention is paid to the size of the truncated basis
$\{ \phi_{nlm}(\mathbf{r}) | 0 \le n \le n_{max}\}$. (In the calculation
we choose for $n_{max}$ such a value for which the classical turning
point of  $\phi_{n_{max}lm}(\mathbf{r})$ is at least 1.5 times bigger
than the
Thomas-Fermi radius. This ensures that we have enough basis functions which
"feel" both regions $\mathbf{r}\in \mbox{TFR}$ and $\mathbf{r} \not\in
\mbox{TFR}$.) In figs.~\ref{fig:energydiff}--\ref{fig:widthscal} we
had $n_{max}=100$ and tested that choosing $n_{max}$ for a
slightly bigger value (cf. $n_{max}$=150) the calculated
$\epsilon_n$, $u_n$ and $v_n$  are practically the same as for $n_{max}=100$.

In fig.~\ref{fig:energydiff}. the energy difference
$E^{(N+1)}-E^{(N)}-\mu$ as given in eq. (\ref{eq:energy_change}) is
plotted. It follows the $N^{1/9}$ law predicted by Son
\cite{Son2007}. Concerning the accuracy of the prefactor, it is
remarkable that even for a relatively small number of particles as
$N=20$ our result lies between the Monte Carlo findings
\cite{Blume2007,Chang2007} and the result of the superfluid density
functional calculation \cite{Bulgac2007}.
(Note that at this
particle number one expects that the extra particle is in the
$s$-state). 
In fig.~\ref{fig:uv}. the components of the eigenfunction are drawn.
They have nodes, the appearence of which can be traced back to the fact that the operator (\ref{eq:tothamrtf}), (\ref{eq:tothamnotrtf})  couples the bare (normal) states of positive and negative eigenvalues.

The Thomas-Fermi approach breaks down in the vicinity of the TF-radius $R$, which circumstance may question the applicability of the pseudopotential (\ref{eq:upsdiagonal}) there. One can show, however, that the width of the solution of eqs. (\ref{eq:modBDE}), (\ref{eq:tothamrtf}), (\ref{eq:tothamnotrtf}) is $\delta r \propto R(d/R)^{4/3}$, where $d$ is the oscillator length $d=\sqrt{\hbar/(m\omega)}$, while the width of the surface region (where gradient corrections are important) scales as $\propto R(d/R)^4$ \cite{Csordas2010}. At large particle numbers $R \gg d$ is valid, which ensures that the error is small when extending the Thomas-Fermi density to the border. 
This makes possible to determine two new universal numbers. In
fig.~\ref{fig:widthscal}. the functions
$w_u=A_u(\hbar\omega/\mu)^{+2/3}$ and
$w_v=A_v(\hbar\omega/\mu)^{+2/3}$ are drawn together with the
numerically determined widths of $\tilde{u}$ and $\tilde{v}$, in units
of $R$. The fitting provides $A_u=1.2774$ and $A_v=1.14616$.

For the solution in case  of $l \ne 0$ one needs a refinement of the model (Then, of course, $\epsilon_{min}$ is to be understood for the given $l$). First of all we quote that at resonance
\begin{equation}
U(\mathbf{r})-\mu=\frac{1}{\xi}\left( U_{ext}(\mathbf{r})-\mu \right),
\end{equation}
where $\xi$ is a universal constant, whose recent values are $0.372
(0.005)$ \cite{Carlson2011} and $0.376(5)$ \cite{Mark2011}. The
complete cancellation of $U(\mathbf{r})-\mu$ by the pseudopotential
occurs now only in a more restricted region, namely in between the
zeros $0<r_1<r_2<R$ of $V(\mathbf{r},\xi)$ defined as 
\begin{equation}
V(\mathbf{r},\xi)=\frac{1}{\xi}\left( U_{ext}(\mathbf{r})-\mu \right)+ \frac{\hbar^2}{2m}\frac{l(l+1)}{r^2}.
\end{equation}
Outside this region, but within the Thomas-Fermi radius $R$ the total potential is $V(\mathbf{r},\xi)$ in the radial equation while outside $R$ it is $V(\mathbf{r},1)$. Semiclassically in the region $r_1<r<r_2$ the radial kinetic energy is positive. Note that in the semiclassical treatment the factor $l(l+1)$ in the centrifugal energy ought to be replaced by $(l+1/2)^2$. We keep the quantum-mechanical expresion, however, to be able to extrapolate the results to small $l$-values.

The solutions of the radial equations corresponding to the potential
discussed above have interesting features. The minimal excitation
energy scales as $\sim\mu^{-1}\sim N^{-1/3}$ as predicted by Son
\cite{Son2007}, but the prefactor depends on the value of
$\xi$ and is smaller than the result 
one gets
from the estimated centrifugal energy at the Thomas-Fermi radius $R$. 
A detailed analysis shows that this effect is mainly due to the two component nature of the wave function. Furthermore, for larger $l$-values the excitation energy is no more proportional to $l(l+1)$. 
At about $l\approx l_{max}$, defined by the condition $r_1=r_2$, the energy curve merges into the excitation spectrum for $l=0$, which might detect a breakdown of the mean-field-type theory at such excitation energies.

One can show that by replacing the wave function of the $(N+1)$-th
atom by the pseudo wave function the wave function of the total system
does not alter. This requirement was the starting point in
\cite{Szepfalusy1955} working within the Hartree-Fock theory (see also
ref.~\cite{Szasz1985} for the generalization to the $(N+2)$-particle
problem). The ground state of the $N$-particle system can be written
as  
\begin{equation}
|\Psi_0\rangle=\prod_t \alpha^{(+)}_t |0\rangle= \prod_t {\alpha^{(-)}_t}^+ |0\rangle,
\label{eq:nbgrstate}
\end{equation}
where ${\alpha^{(-)}}^+$ and ${\alpha^{(+)}}^+$ are the quasiparticle creation operators in negative and positive energy states, respectively, with $\alpha^{(\pm)}$ the corresponding destruction operators. $|0\rangle$ stands for the vacuum of the atoms.

The state of the $(N+1)$-body system can be given as a one-quasiparticle state
\begin{equation}
\alpha^+_n | \Psi_0 \rangle.
\label{eq:npobgrstate}
\end{equation}
The corresponding state written in terms of the pseudo wave function reads as
\begin{equation}
\left(\alpha^+_n+\sum_t c_t^{(+)}\alpha_t^{(+)}+\sum_t {c_t^{(-)}}^* {\alpha_t^{(-)}}^+\right)| \Psi_0 \rangle,
\label{eq:npobgrstatediffform}
\end{equation}
which is equivalent to (\ref{eq:npobgrstate}) as can be seen from the expression of $| \Psi_0 \rangle $. Note that normalization constants have not been included in the discussion above. The coefficients in (\ref{eq:npobgrstatediffform}) are 
\begin{equation}
c_t^{(\pm)}=\left(\epsilon_n \pm \epsilon_t \right) \langle \phi_t^{(\pm)}|\tilde{\phi}_n \rangle.
\end{equation}

In conclusion we have calculated the energy of an extra particle in
trapped Fermi gases at unitarity by generalizing the pseudopotential
theory to the BdG-type equation. The background $N$-particle problem has been
treated within the Thomas-Fermi theory. This approximation could be
improved by including the Weizs\"acker-type correction as determined
in case of the unitary system \cite{Csordas2010}. Along with such an
extension it would be appropriate to take into account also the
nonlocality of the pseudopotential, a feature lost when obtaining
(\ref{eq:upsmatrix}). All these are left for future work.   

Some final remarks are in order. In our treatment all about
the neighbouring even-number state are compressed into two universal
numbers $\delta$ and $\xi$. The recent value of $\xi$ is somewhat
smaller than previously used in different matching processes, which
might influences the other parameter values. This makes
our choice to take the bare atomic mass in our
calculation reasonable. Concerning $\delta$ we have taken its mean-field value
by the same reason and by the expectation that it alters only
slightly when using different models.

Applications of more elaborated forms of the  pseudopotential
to other systems along
with details of the present work will be published elsewhere.

The present work has been partially supported by the Hungarian Scientific
Research Fund under Grant Nos. OTKA 77534/77629 and OTKA 75529.


\bibliographystyle{eplbib}
\bibliography{pseudopot}

\end{document}